\documentclass{JHEP3}
\usepackage{graphicx}
\usepackage{epsfig}
\usepackage{amssymb}
\usepackage{amsmath}
\usepackage{float}

\newcommand{\bea}{\begin{eqnarray}}
\newcommand{\eea}{\end{eqnarray}}
\newcommand{\bean}{\begin{eqnarray*}}
\newcommand{\eean}{\end{eqnarray*}}
\newcommand{\nn}{\nonumber \\}

\def\WH #1{\widehat{#1}}

\def\a{{\alpha}}

\def\b{{\beta}}

\def\Label#1{\label{#1}%
  \smash{\hbox to0pt{\raise1ex\hbox{\tiny[#1]}\hss}}}

\title{Note on Permutation Sum of Color-ordered Gluon Amplitudes }

\author{Yi-Jian Du${}^{a}$, Bo Feng${}^{a,b}$, Chih-Hao Fu${}^{b}$    \\
$^a$\small Zhejiang Institute of Modern Physics, Zhejiang
University, Hangzhou, 310027, P. R. China\\$^b$\small Center of
Mathematical Science, Zhejiang University, Hangzhou, 310027, P. R. China \\
}

\date{\today}
\abstract{In this note we show that under BCFW-deformation the
large-$z$ behavior of permutation sum of color-ordered gluon
amplitudes found  by Boels and Isermann in arxiv:1109.5888 can be
simply understood from the well known Kleiss-Kuijf relation and
Bern-Carrasco-Johansson relation. }


\begin{document}

\section{Introduction}

Initiated by Witten's work on twistor theory \cite{Witten:2003nn},
a tremendous amount of progress has been made in the calculation
and understanding
of scattering amplitudes. One of such  is  the on-shell
recursion relation (BCFW recursion relation) for tree-level gluon
amplitudes \cite{Britto:2004ap, Britto:2005fq}. The study of
recursion relation originated from analyzing BCFW-deformation on a
particular pair of particles $(i,j)$
\bea  p_i\to p_i-zq, ~~~p_j\to p_j+z q,~~~q^2=q\cdot p_i=q\cdot
p_j=0~.~~~\Label{BCFW-def}\eea
Under the deformation, the physical on-shell amplitude $A_n$ of
$n$-particles becomes a rational function $A_n(z)$ of single
complex variable $z$ with only pole structures\footnote{It can be
single poles at finite values of $z$ or multiple pole at the
$z=\infty$.}. The large-$z$ behavior (or  the ``boundary behavior'')
of $A_n(z)$ under such a deformation turns out to be very important
for our deep understanding of many properties in  quantum field
theories. The determinant of boundary behavior is not so easy and a
naive analysis from Feynman diagrams could often lead to wrong
conclusions. A very nice  analysis was done by Arkani-hamed and Kaplan in
\cite{ArkaniHamed:2008yf}, where because  $zq\to \infty$, whole
amplitude can be considered as scattering of particles $i,j$ from
soft background constructed by other particles.

The boundary behavior can be classified into two categories. The first
category is that $A(z)$ does not vanish when $z\to \infty$. For this
category, to write down on-shell recursion relation, we need to find
 boundary contributions. So far there is no general theory to
extract such information easily, however some progress can be found in
\cite{Feng:2009ei,Feng:2010ku,Benincasa:2011kn,Feng:2011tw,Benincasa:2011pg}.
The second category is that $A(z)\to 0$ when $z\to \infty$. For this
case, the boundary contribution is zero and the familiar on-shell
recursion relation is derived based on this condition.

The vanishing behavior of second category is a little bit rough and
we can make it more accurate by writing the leading behavior as
$A(z)\sim {1\over z^k},~k\geq 1$. If for some theories we have
$k\geq 2$, more relations can be derived in addition to the standard BCFW recursion
relation. These ``bonus'' relations were discussed in
\cite{Benincasa:2007qj, ArkaniHamed:2008yf,
Spradlin:2008bu,Badger:2008rn, Feng:2010my,Badger:2010eq,
He:2010ab}, where their usefulness was demonstrated from
various aspects. In  particular, the better vanishing behavior of
color-ordered gluon amplitudes with non-adjacent deformation pair was
used in \cite{Feng:2010my} to prove 
 the BCJ relation \cite{Bern:2008qj}\footnote{The BCJ relation has first
been proved in string theory
\cite{BjerrumBohr:2009rd,Stieberger:2009hq}, and then in field
theory \cite{Feng:2010my, Chen:2011jxa}.} between color-ordered
gluon amplitudes. An important consequence of BCJ relation is that
among all $(n-1)!$ color-ordered amplitudes, only $(n-3)!$ of them
are needed and others can be  written as the linear combinations of
the basis.

Recently, another new better vanishing behavior under deformation
(\ref{BCFW-def}) was observed by Boels and Isermann in
\cite{Boels:2011tp}, which can be summarized as the following two
statements\footnote{Incidentally a second paper of this series of work
came out during the preparation of this note, where similar argument
was used \cite{Boels:2011mn}.}
\bea \sum_{perm~\a} A_n (i, \{\a\}, j, \{\b\})\to
\xi_{i\mu}(z)\xi_{j\nu}(z){ G^{\mu\nu}(z)\over z^{k}},~~~k=\left\{
\begin{array}{ll} n_\a, &  i,j~not~nearby \\ n_\a-1,~~~ &  i,j~~nearby
\end{array} \right.~~~\Label{Boels-1}\eea
and
\bea \sum_{cyclic~\a} A_n (i, \{\a\}, j, \{\b\})\to
\xi_{i\mu}(z)\xi_{j\nu}(z){ G^{\mu\nu}(z)\over z^{k}},~~~k=\left\{
\begin{array}{ll} 2, &  i,j~not~nearby \\ 1,~~~ &  i,j~~nearby
\end{array} \right.~~~\Label{Boels-2}\eea
where $n_\a$ is the number of elements in set $\a$, $\xi$ is the
polarization vector and $G_{\mu \nu}$ is given by
\cite{ArkaniHamed:2008yf}
\bea G_{\mu \nu}= z\eta^{\mu\nu} f(1/z)+ B^{\mu\nu}(1/z)+{\cal
O}(1/z)~.\eea
With these new results, a natural question to ask is that do they provide
new nontrivial relations among color-ordered amplitudes or they can
be understood from known results?

In this short note we would like to answer the above question. In
particular we shall show that using the familiar KK-relation and
fundamental BCJ relation, (\ref{Boels-1}) can be easily understood.
 The  Kleiss-Kuijf (KK) relation  was first conjectured in
\cite{Kleiss:1988ne} and proved in \cite{DelDuca:1999rs}. The
formula reads
\bea  A_n(1,\{\a\}, n,\{\b\}) = (-1)^{n_\b}\sum_{\sigma\in
OP(\{\a\},\{\b^T\})} A_n(1,\sigma, n)~,~~~~\Label{KK-rel}\eea
where the Order-Preserved (OP) sum is to be taken over all permutations
of set $\a \bigcup \b^T$ whereas the relative ordering in  sets  $\a$  and
$\b^T$ (, which is the reversed ordering of set $\b$) are preserved.
The $n_\b$ here is the number of elements in set $\b$ . One
non-trivial example with six gluons is given as the following
\bea A(1,\{ 2,3\},6,\{4,5\}) & = & A(1,2,3,5,4,6)+ A(1,2,5,3,4,6)+
A(1,2,5,4,3,6)\nn & & + A(1,5,4,2,3,6)+A(1,5,2,4,3,6)+A(1,5,2,3,4,6)
~~~\Label{KK-6-point}\eea
Comparing to the simple expression of KK-relation (\ref{KK-rel}),
the general BCJ relation given in \cite{Bern:2008qj} is complicated,
where the basis with three particles  fixed at three positions has
been used to expand other amplitudes.  However, there is a very
simple relation, which we will call the ``fundamental BCJ relation''.
The fundamental BCJ relation  can be used to derive other BCJ
relations and its  expression  is given by
\bea 0= s_{21}
A_n(1,2,3,...,n-1,n)+\sum_{j=3}^{n-1}(s_{21}+\sum_{t=3}^j s_{2t})
A(1,3,4,..,j,2,j+1,...,n)~.~~~~\Label{Fund-BCJ}\eea

The plan of the note is following. In section two we prove the
conjecture (\ref{Boels-1}) for $i,j$ not nearby. The nearby case is
special and the proof is given in section three. Finally, a brief
summation and discussion are given in section four.

\section{The non-adjacent case}

Without lost of generality we pick the shifted pair to be
$(1,j)$ and define the following sum
\bea T_j(z)\equiv \sum_{\sigma\in S_{j-2}} A_n(\WH
1(z),\sigma(2),...,\sigma(j-1), \WH j(z),...n-1,n),~~~~j=2,...,n-1~.
~~~~\Label{Sum-nonearby}\eea
where the case $j=n$ is the special adjacent case to be
discussed in next section. In the following we will encounter
cases where permutation sums are  performed on  $(j-2)$ elements other than
$(2,...,j-1)$ in (\ref{Sum-nonearby}). For simplicity we will stick to the
same notation $T_j$, but the accurate definitions of these $T_j$
should be clear from our discussions.

According to the conjecture (\ref{Boels-1}), we should have
\bea \lim_{z\to\infty} T_j(z)\to \xi_{1\mu}(z) \xi_{j\nu}(z)
{G_{\mu\nu}(z)\over z^{j-2}}~.~~~~\Label{Tj-z}\eea
For the case $j=2$ and $j=3$ we have
\bea \lim_{z\to\infty} T_2(z)&= & \lim_{z\to\infty} A_n(\WH 1(z),\WH
2(z),3,...n-1,n)\to \xi_{1\mu}(z) \xi_{j\nu}(z) {G_{\mu\nu}(z)}\nn
\lim_{z\to\infty} T_3(z)& =& \lim_{z\to\infty} A_n(\WH 1(z),2, \WH
3(z),4,...n-1,n)\to \xi_{1\mu}(z) \xi_{j\nu}(z) {G_{\mu\nu}(z)\over
z}~~~\Label{T2-T3}\eea
which are well known to be true from
\cite{ArkaniHamed:2008yf}\footnote{In fact assuming the result is
true for $j=2$, the case with $j=3$ can be proved using fundamental BCJ
relation as is done in following paragraphs. To avoid arguing in a
circle we need to establish the BCJ relation from, for example,
string theory \cite{BjerrumBohr:2009rd,Stieberger:2009hq},  in stead
of  from the bonus relation given in \cite{Feng:2010my}.}.

For the case $j=4$, let us consider following two fundamental BCJ
relations
\bea 0 & = & B_{23}= s_{2\WH 1} A_n(\WH 1(z),2,3,\WH 4(z),5,...,n) +
(s_{2\WH 1}+s_{23}) A_n(\WH 1(z),3,2,\WH 4(z),5,...,n)\nn & &
+\sum_{k=4}^{n-1}(\sum_{t=1}^k s_{2k}) A_n(\WH 1,3,\WH
4,...k,2,k+1,...,n) \eea
and its corresponding $B_{32}$ by exchanging $2\leftrightarrow 3$.
To emphasize the $z$-dependence, we have put on a ``hat'' for clarity,
thus $s_{2\WH 1}\neq s_{21}$.  One key observation is that the third
term in $B_{23}$ is independent of $z$ because $s_{2\WH 1}+s_{2\WH
4}=s_{21}+s_{24}$. Summing them up, we have
\bean  0 & = & B_{23}+B_{32}= (s_{2\WH 1}+s_{3\WH 1}+s_{2 3}) T_{4}
+ \left(\sum_{k=4}^{n-1}(\sum_{t=1}^k s_{2k}) A_n(\WH 1,3,\WH
4(z),...k,2,k+1,...,n)+\left\{2\leftrightarrow 3\right\}\right)\eean
or
\bea T_4 & = & {\left(\sum_{k=4}^{n-1}(\sum_{t=1}^k s_{2k}) A_n(\WH
1,3,\WH 4(z),...k,2,k+1,...,n)+\left\{2\leftrightarrow
3\right\}\right)\over -s_{\WH 1 23}}~.~~~\Label{T4-BCJ}\eea
Since each term in the numerator of (\ref{T4-BCJ}) is the form of
$T_3$ defined in (\ref{Sum-nonearby}), using (\ref{T2-T3}) we have
immediately
\bea \lim_{z\to\infty} T_4(z)\to {1\over z} \lim_{z\to\infty}
T_3(z)\to  \xi_{1\mu}(z) \xi_{j\nu}(z) {G_{\mu\nu}(z)\over
z^{2}}~~~\Label{T4}\eea
which is the result we want.

To see more clearly our method, let us consider the case $j=5$. For
this we write down the following BCJ relations
\bea B_{234} & = & (s_{2\WH 1}) A(\WH 1,2,3,4,\WH 5,...) + (s_{2\WH
1}+s_{23}) A(\WH 1,3,2,4,\WH 5,...)+ (s_{2\WH 1}+s_{23}+s_{24})
A(\WH 1,3,4,2,\WH 5,...)\nn & & + ...+(\sum_{t=1}^js_{2t})A(\WH
1,3,4,\WH 5,...,j,2,j+1,...)\nn
B_{243} & = & (s_{2\WH 1}) A(\WH 1,2,4,3,\WH 5,...) + (s_{2\WH
1}+s_{24}) A(\WH 1,4,2,3,\WH 5,...)+ (s_{2\WH 1}+s_{23}+s_{24})
A(\WH 1,4,4,2,\WH 5,...)\nn & & + ...+(\sum_{t=1}^js_{2t})A(\WH
1,4,3,\WH 5,...,j,2,j+1,...)\eea
and the corresponding $B_{324}, B_{342}, B_{423}, B_{432}$. The first
observation is that from the sum $B_{234}+B_{243}$,
\bean (\sum_{t=1}^js_{2t})A(\WH 1,3,4,\WH
5,...,j,2,j+1,...)+(\sum_{t=1}^j s_{2t})A(\WH 1,4,3,\WH
5,...,j,2,j+1,...)\sim  (\sum_{t=1}^js_{2t})T_{j=4}~.\eean
A second observation is that from the sum $(B_{234}+{\rm perm})$
we get
\bean & & (s_{2\WH 1}+s_{3\WH 1}+s_{4\WH 1}+s_{23}+s_{24}+s_{34})
\sum_{\sigma\in S_3(2,3,4)} A(\WH 1,
\sigma(2),\sigma(3),\sigma(4),\WH 5,...)=s_{\WH 1234} T_{j=5}~.\eean
Now using $(B_{234}+{\rm perm})=0$ we can solve
\bea T_{j=5}= {\sum_{a} c_a T_{j=4}^a\over -s_{\WH 1 234}}\eea
where $T_{j=4}^a$ are various combinations of type $T_{j=4}$ and
$c_a$ are $z$-independent kinematic coefficients. Using the boundary
behavior of type $T_{j=4}$ (\ref{T4}) we have immediately
\bea \lim_{z\to\infty} T_5(z)\to {1\over z} \lim_{z\to\infty}
T_4(z)\to \xi_{1\mu}(z) \xi_{j\nu}(z) {G_{\mu\nu}(z)\over
z^{3}}~.\eea

Having seen above two examples, we give the general proof. Assuming
(\ref{Tj-z}) is true for $j=k-1$, we will show it is true for $j=k$.
To do so, we consider the sum $(B_{23...(k-1)}+{\rm
perm(2,...,k-1)})$ where $B_{23...(k-1)}$ is following fundamental
BCJ relation
\bean 0 & = & B_{23...(k-1)}= \sum_{i=2}^{k-1}(s_{2\WH
1}+\sum_{t=2}^i s_{2t}) A_n(\WH 1,3,..,i,2,i+1,...,\WH k,...,n)\nn &
& + \sum_{i=k}^{n-1}(s_{2 1}+\sum_{t=2}^i s_{2t}) A_n(\WH
1,3,4,...,\WH k, ...,i, 2,i+1,...,n)~.\eean
From the sum we can solve
\bea  T_{j=k}= { \sum_{a,b} c_{a,b} T_{j=k-1}^{a,b}\over -s_{\WH 1 2
3 ...(k-1)}}~,\eea
where $c_{a,b}  =  \sum_{t=1}^{b} s_{at}$ and
\bean T_{k-1}^{a,b}(z)& \equiv & \sum_{\sigma\in S_{k-3}(2,3,..,a-1,
a+1,...,k-1)} A_n(\WH 1(z),\{\sigma\}, \WH k(z),...,b,a,b+1,n-1,n),
\eean
thus we have
\bea \lim_{z\to\infty} T_k(z)\to {1\over z} \lim_{z\to\infty}
T_{k-1}(z)\to \xi_{1\mu}(z) \xi_{j\nu}(z) {G_{\mu\nu}(z)\over
z^{k-2}}~.\eea

It is obvious that above  method will not work for the adjacent $T_{j=n}$
because the factor $s_{\WH 12...(n-1)}=s_{\WH n}=0$.

\section{The adjacent case}

Having proved the conjecture (\ref{Boels-1}) for non-adjacent case,
we move on to the special adjacent case. It can happen when and only
when $i,j$ are at the two ends and  permutation sum is  over all
remaining $(n-2)$ elements. Without losing generality, we fix
$(i,j)$ to be $(1,n)$, thus we have
\bea T_{n}=\sum_{\sigma\in S_{n-2}} A_n(\WH 1(z),
\sigma(2),...,\sigma(n-1), \WH n(z))~,~~~~\Label{ij-nearby}\eea
 where again to emphasize the
$z$-dependence we put  hats on particles $(1,n)$. Before giving
general argument, let us see the example $n=5$. There are six terms
in the sum and we can group them as
\bean T_5 & = &  \left[ A(\WH 1, 2,3,4,\WH 5)+ A(\WH 1, 3,2,4,\WH
5)+A(\WH 1, 3,4,2,\WH 5)\right]\nn & & + \left[ A(\WH 1,2,4,3,\WH
5)+A(\WH 1,4,2,3,\WH 5)+A(\WH 1,4,3,2,\WH 5)\right]~.\eean
For each group, using the KK-relation (\ref{KK-rel}), we can write
them as
\bea T_5 & = & - A(\WH 1,3,4,\WH 5, 2)- A(\WH 1,4,3,\WH 5, 2)\eea
where the right hand side is nothing but the non-nearby type
$T_{j=4}$. A byproduct is that all $T_{j=4}$ are identical, no matter
which element of $(2,3,4)$ was put at the  rightmost position.

Having the example of $n=5$, we explain how to regroup  terms in the
sum (\ref{ij-nearby}). These $(n-2)!$ terms can be divided into
$(n-2)$ groups $G_k$, where for each group particle $2$ is at the
fixed position $k$ with $k=2,...,n-1$. Now considering a given
ordering $\sigma(3),...,\sigma(n-1)$ determined by permutations over
remaining $(n-3)$ elements, in each group there is
 an unique amplitude with the given ordering $\sigma(3),...,\sigma(n-1)$
 and we sum them up  to get
\bea I_{\sigma} &= & A_n (\WH 1(z),2, \sigma(3),...,\sigma(n-1), \WH
n(z))+A_n (\WH 1(z), \sigma(3),2,\sigma(4),...,\sigma(n-1), \WH
n(z)) \nn &  & +...+A_n (\WH 1(z), \sigma(3),...,2,\sigma(n-1), \WH
n(z))+A_n (\WH 1(z), \sigma(3),\sigma(4),...,\sigma(n-1), 2,\WH
n(z))~~~~\Label{ij-sigma}\eea
There are $(n-2)$ terms in (\ref{ij-sigma}) and the sum is nothing
but
\bea I_\sigma= -A_n (\WH 1(z), \sigma(3),...,\sigma(n-1), \WH n(z),
2) \eea
by the familiar KK-relation (\ref{KK-rel}). Thus we have
\bea T_{n} & = & \sum_{\sigma \in S_{n-3}} I_\sigma= -\sum_{\sigma
\in S_{n-3}} A_n (\WH 1(z), \sigma(3),...,\sigma(n-1), \WH n(z),
2)=-T_{n-1}~,\eea
and  from (\ref{Tj-z}) in previous section we have
\bea \lim_{z\to\infty} T_n(z)=- \lim_{z\to\infty} T_{n-1}(z)\to
\xi_{1\mu}(z) \xi_{j\nu}(z) {G_{\mu\nu}(z)\over z^{n-3}}~\eea
which is the conjecture given in (\ref{Boels-1}). Again, from our
derivation  all $T_{j=n-1}$ are identical as long as the deformation pair
$(1,n)$ is same.

\section{Conclusion}

Recently there was a new  conjecture proposed by Boels and Isermann
 in \cite{Boels:2011tp},
which stated  that under the BCFW-deformation the large-$z$ behavior
of partial permutation sum of color-ordered gluon amplitudes has
better convergent behavior. From our previous experiences better
convergent behavior could lead to new ``bonus'' relations among
color-ordered gluon amplitudes. However, either from string theory
or from BCJ relation, the minimum basis of $n$-point
amplitudes was known to consist of $(n-3)!$ independent amplitudes,
thus there is a potential contradiction between
new result and our old intuition picture.

In this short note, using the KK-relation and fundamental BCJ
relation, we presented a simple proof of the conjecture made in
\cite{Boels:2011tp}. Our proof demonstrated that the better convergent
behavior is a natural consequence of known results, so the ``naive''
contradiction does not exist.

Besides the conjecture discussed in this note, there are other
combinations having better convergent behaviors  as given in
\cite{Boels:2011tp}. We feel that it would be interesting to investigate them
from the new aspect. Finally working out full consequences of these
results in tree and loop amplitudes would be also very important. For
this direction,  some progress has been made in
\cite{Boels:2011tp}.

\subsection*{Acknowledgements}
Y. J. Du is supported in part by the NSF of China Grant No.11105118,
No.11075138. B. Feng is supported by fund from Qiu-Shi, the
Fundamental Research Funds for the Central Universities with
contract number 2010QNA3015, as well as Chinese NSF funding under
contract No.10875104, No.11031005, No.11135006, No. 11125523. CF
would like to thank Yu-tin Huang for valuable discussions and the
2011 Simons workshop in Mathematics and Physics for hospitality.


\end{document}